\documentclass[prl,superscriptaddress,twocolumn,nopreprintnumbers,
floatfix,preprintnumbers,altaffilletter]{revtex4}
\usepackage{graphicx,url,amssymb,amsmath,longtable,rotating,color,units,
wasysym,subfigure,epsfig,multirow}

\newcommand{\msun}{M_\odot}

\begin{document}

\title{The promise of multi-band gravitational wave astronomy after GW150914}

\author{Alberto Sesana}
\affiliation{ School of Physics and Astronomy, University of
Birmingham, Edgbaston, Birmingham B15 2TT, United Kingdom}

\begin{abstract}
The black hole binary (BHB) coalescence rates inferred from the advanced LIGO (aLIGO) detection of GW150914 imply an unexpectedly loud GW sky at milli-Hz frequencies accessible to the evolving Laser Interferometer Space Antenna (eLISA), with several outstanding consequences. First, up to thousands of BHB will be individually resolvable by eLISA; second, millions of non resolvable BHBs will build a confusion noise detectable with signal-to-noise ratio of few to hundreds; third -- and perhaps most importantly -- up to hundreds of BHBs individually resolvable by eLISA will coalesce in the aLIGO band within ten years. eLISA observations will tell aLIGO and all electromagnetic probes weeks in advance when and where these BHB coalescences will occur, with uncertainties of $<10$s and $<1$deg$^2$. This will allow the pre-pointing of telescopes to realize coincident GW and multi-wavelength electromagnetic observations of BHB mergers. Time coincidence is critical because prompt emission associated to a BHB merger will likely have a duration comparable to the dynamical time-scale of the systems, and is only possible with low frequency GW alerts.
\end{abstract}

\maketitle

{\it Introduction.} The two aLIGO \cite{2010CQGra..27h4006H} detectors observed a black hole binary (BHB) of 36$^{+5}_{-4}\msun$ and 29 $^{+4}_{-4}\msun$ coalescing at $z=0.09^{+0.03}_{-0.04}$ on September 14, 2015 (GW150914) \cite{2016PhRvL.116f1102A}. This observation has been used to observationally constrain the cosmic merger rate of BHB for the first time \cite{2016arXiv160203842A}. Although theoretical predictions vary wildly \cite{2010CQGra..27q3001A}, spanning the whole range constrained by the limit implied by initial LIGO \cite{2013PhRvD..87b2002A}, high rates of several hundreds event yr$^{-1}$Gpc$^{-3}$ have been suggested by some authors \cite{2010ApJ...715L.138B,2015ApJ...806..263D}. GW150914 sets the bar in this range, finding rates between 2 and 400 yr$^{-1}$Gpc$^{-3}$ (depending on the assumed BHB mass distribution). Moreover, the observed system is three times heavier than the 'canonical' $10\msun+10\msun$ binary (but see \cite{2015arXiv151004615B,2014MNRAS.442.2963K}), implying a gravitational wave (GW) strain amplitude a factor 3$^{5/3}$ larger.

\begin{figure}[ht]
\begin{center}
\includegraphics[scale=0.43,clip=true,angle=0]{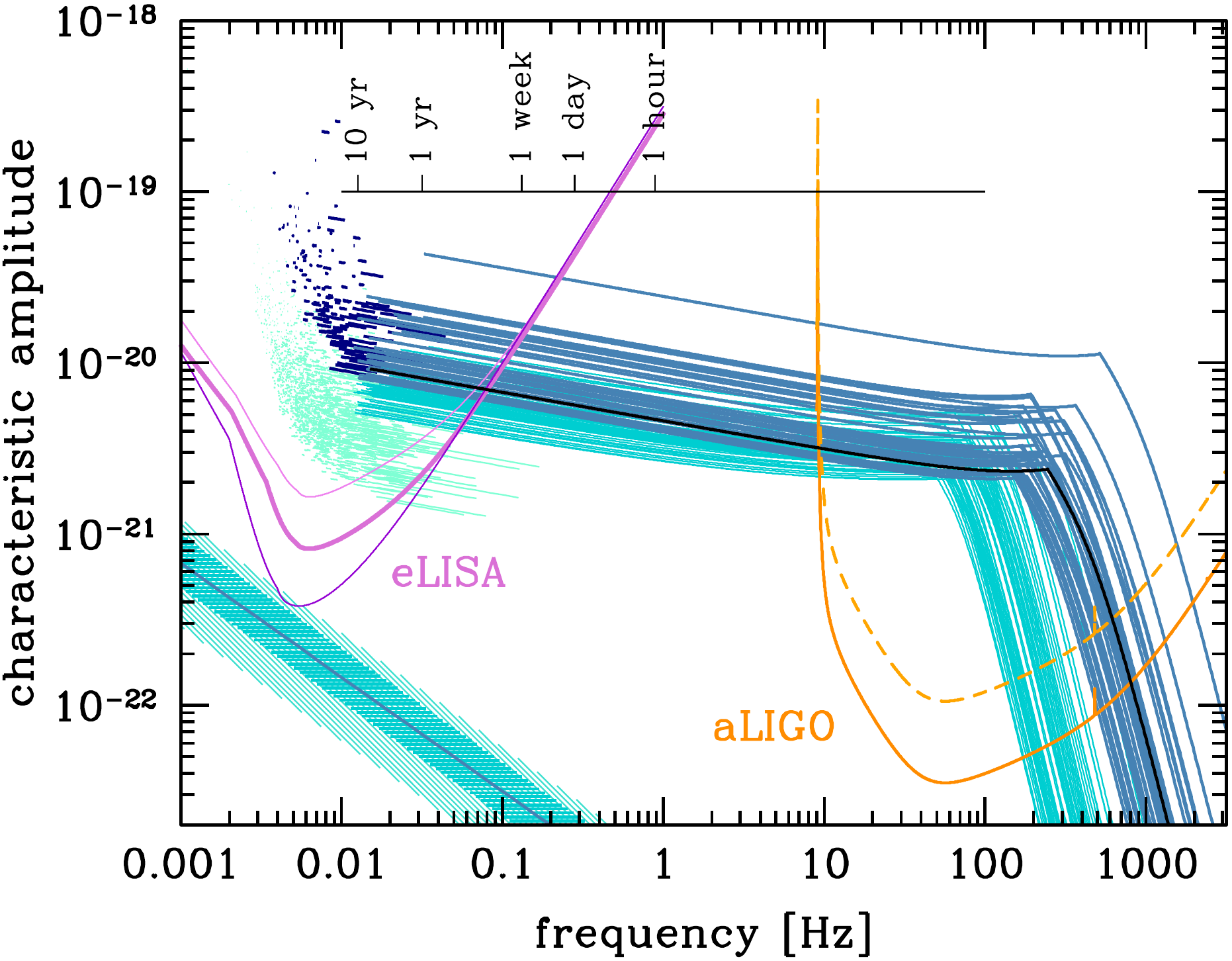}
  \caption{The multi-band GW astronomy concept. The violet lines are the total sensitivity curves (assuming two Michelson) of three eLISA configurations; from top to bottom N2A1, N2A2, N2A5 (from \cite{2016PhRvD..93b4003K}). The orange lines are the current (dashed) and design (solid) aLIGO sensitivity curves. The lines in different blue flavours represent characteristic amplitude tracks of BHB sources for a realization of the {\it flat} population model (see main text) seen with S/N$>1$ in the N2A2 configuration (highlighted as the thick eLISA middle curve), integrated assuming a five year mission lifetime. The light turquoise lines clustering around 0.01Hz are sources seen in eLISA with S/N$<5$ (for clarity, we down-sampled them by a factor of 20 and we removed sources extending to the aLIGO band); the light and dark blue curves crossing to the aLIGO band are sources with S/N$>5$ and S/N$>8$ respectively in eLISA; the dark blue marks in the upper left corner are other sources with S/N$>8$ in eLISA but not crossing to the aLIGO band within the mission lifetime. For comparison, the characteristic amplitude track completed by GW150914 is shown as a black solid line, and the chart at the top of the figure indicates the frequency progression of this particular source in the last 10 years before coalescence. The shaded area at the bottom left marks the expected confusion noise level produced by the same population model (median, 68\% and 95\% intervals are shown). The waveforms shown are second order post-Newtonian inspirals phenomenologically adjusted with a Lorentzian function to describe the ringdown.}
\label{fig1}
\end{center}
\end{figure}

Assuming a circular binary, GW150914 was emitting at a frequency of $\approx0.016$ Hz five years prior to coalescence, well within the eLISA \cite{2013arXiv1305.5720C} band (see figure \ref{fig1}). Even more remarkably, the signal-to-noise (S/N) accumulated in an eLISA type detector in those final years (sky and polarization averaged) would have varied between three and fifteen, depending on the detector configuration (see below): had eLISA been operating, we would have known exactly when and where GW150914 would have appeared in the aLIGO data. This opens the prospect of multi-band GW astronomy, as illustrated in figure \ref{fig1}; BHBs emit in the eLISA band for years before eventually chirping to high frequency producing a short (but strong) signal in aLIGO. Multi-band GW astronomy has already been proposed in the context of observing either population III merger remnants \cite{2009ApJ...698L.129S} or intermediate mass BHB \cite{2010ApJ...722.1197A,2012PhRvD..85l3005K} with LISA and aLIGO/Einstein Telescope \cite{2010CQGra..27s4002P}. However, with GW150914 detection and the inferred rates there are a number of profound consequences for aLIGO, eLISA and electromagnetic follow-ups to LIGO sources, that we describe in this Letter.

{\it BHB population models.} Based on the observation of GW150914, the probability distribution of the intrinsic comoving merger rate ${\cal R}$ under two distinct assumptions for the BHB mass function was computed in \cite{2016arXiv160203842A}. In model {\it flat}, the masses of the two BHs, $M_{1,r}$ \& $M_{2,r}$ (the subscript $_r$ refers to quantities measured in the rest frame of the source) are independently drawn from a log-flat distribution in the range $5\msun<M_{1,2,r}<100\msun$, with the restriction of BHB total mass being in the range $5\msun - 100\msun$. In model {\it salp}, $M_{1,r}$ is drawn from a Salpeter mass function in the range $5\msun<M_{1,2,r}<100\msun$ and $M_{2,r}$ from a flat distribution between $5\msun$ and $M_{1,r}$. The {\it flat} model implies a heavy-biased BHB mass function, with a characteristic BHB merger rate $\approx 35$yr$^{-1}$Gpc$^{-3}$, whereas the {\it salp} model favours relatively light BHBs, compensated by a higher rate ($\approx 100$yr$^{-1}$Gpc$^{-3}$), for consistency with the observation of GW150914. We assume no cosmic evolution of this intrinsic rate.

For each of the mass models we compute $p({\cal M}_r)$ -- the associated chirp mass probability distribution, where ${\cal M}_r=(M_{1,r}M_{2,r})^{3/5}/(M_{1,r}+M_{2,r})^{1/5}$, and we multiply it by the comoving merger rate ${\cal R}$, thus obtaining the merger rate density per unit mass 
\begin{equation}
  \frac{d^2n}{d{\cal M}_rdt_r}={\cal R}\times p({\cal M}_r).
  \label{eqrate}
\end{equation}
Equation (\ref{eqrate}) can be then converted into a number of sources emitting  per unit mass, redshift and frequency at any time via
\begin{equation}
  \frac{d^3N}{d{\cal M}_rdzdf_r}=\frac{d^2n}{d{\cal M}_rdt_r}\frac{dV}{dz}\frac{dt_r}{df_r},
  \label{eq1}
\end{equation}
where $dV/dz$ is the standard volume shell per unit redshift in the fiducial $\Lambda$CDM cosmology ($h=0.679,\,\Omega_M=0.306,\,\Omega_\Lambda=0.694$, \cite{2015arXiv150201589P}), and $dt_r/df_r$ is given by
\begin{equation}
  \frac{dt_r}{df_r}=\frac{5c^5}{96\pi^{8/3}}(G{\cal M}_r)^{-5/3}f_r^{-11/3}.
  \label{eq2}
\end{equation}
Equation (\ref{eq2}) is valid for circular binaries, which is our working hypothesis. This is certainly a good approximation for systems formed through stellar evolution, that are expected to inherit their stellar progenitor circular orbits. Extrapolating results shown in figure 10 of \cite{2016arXiv160202444R} at low frequency, we find that also dynamically formed BHBs have typical $e \lesssim 0.01$ in the relevant eLISA band, making our S/N and source number computations robust against the assumed BHB formation channel.

For both the {\it flat} and {\it salp} models, probability distributions of the intrinsic rate ${\cal R}$ are given in \cite{2016arXiv160203842A} (see their figure 5). We make 200 Monte Carlo draws from each of those, use equation (\ref{eq1}) to numerically construct the cosmological distribution of emitting sources as a function of mass redshift and frequency, and make a further Monte Carlo draw from the latter. For each BHB mass model, the process yields 200 different realizations of the instantaneous BHB population emitting GWs in the Universe. We limit our investigation to $0<z<2$ and $f_r>10^{-4}$Hz, sufficient to cover all the relevant sources emitting in the eLISA and aLIGO bands. 

{\it Signal-to-noise ratio computation.} An in-depth study of possible eLISA baselines in under investigation \cite{2016PhRvD..93b4003K,2015arXiv151206239C,2016arXiv160107112T}, and the novel piece of information we provide here might prove critical in the selection of the final design. Therefore, following \cite{2016PhRvD..93b4003K}, we consider six baselines featuring one two or five million km arm-length (A1, A2, A5) and two possible low frequency noises -- namely the LISA Pathfinder goal (N1) and the original LISA requirement (N2). We assume a two Michelson (six laser links) configuration, commenting on the effect of dropping one arm (going to four links) on the results. We assume a five year mission duration.

In the detector frame, each source is characterized by its {\it redshifted} quantities ${\cal M}={\cal M}_r(1+z)$ and $f=f_r/(1+z)$. During the five years of eLISA observations, the binary emits GWs shifting upwards in frequency from an initial value $f_i$, to an $f_f$ that can be computed by integrating equation (\ref{eq2}) for a time $t_r=5{\rm yr}/(1+z)$. The sky and polarization averaged S/N in the eLISA detector is then computed as
\begin{equation}
  ({\rm S/N})^2 = 2\int_{f_i}^{f_{\rm f}}\frac{h_c^2(f)}{f\langle S(f)\rangle}d{\rm ln}f,
\label{eqsnr}
\end{equation}
where the factor 2 accounts for the fact that we have two Michelson interferometers (i.e. we consider six laser links). $h_c$ is the characteristic strain of the source given by
\begin{equation}
 h_c= \frac{1}{\pi D}\left(\frac{2G}{c^3}\frac{dE}{df}\right)^{1/2},
\label{hc}
\end{equation}
where $D$ is the comoving source distance, and the emitted energy per unit frequency is
\begin{equation}
\frac{dE}{df}=\frac{\pi}{3G}\frac{(G{\cal M})^{5/3}}{1+z}(\pi f)^{-1/3}.
\label{hc}
\end{equation}
In equation (\ref{eqsnr}), $\langle S(f)\rangle$ is the eLISA instrumental noise, averaged over the source sky location and wave polarization, and it is estimated by using the analytical form given in \cite{2016PhRvD..93b4003K} for each configuration. Note that at the high frequencies relevant for the sources crossing to the aLIGO band, the real eLISA sensitivity is not well captured by the analytical fitting functions. However, this does not appreciably affect S/N computations, and is not expected to significantly alter detector performances (Petiteau et al. in preparation). For parameter estimation, we adopt a modification of the Fisher Matrix code of \cite{2005PhRvD..71h4025B}. The code employs a 3.5 Post Newtonian (3.5PN) circular non-spinning gravitational waveform evaluated in the frequency domain assuming the stationary phase approximation. The limitation to non-spinning, circular binaries is not critical here, since the main source parameters of interest are the sky localization and the time to coalescence. The former depends mostly on the signal Doppler modulation and the time-varying antenna beam pattern due to the detector's orbital motion, neither of which is influenced by the adopted waveform. The latter mostly depends on the estimate of the redshifted chirp mass, which is automatically determined to $\approx 10^{-6}$ (see figure \ref{fig3}) relative precision by match filtering hundreds of thousand of source cycles. In fact, preliminary results using 3.5PN spinning precessing waveforms confirm the figures shown in the following (A. Klein, private communication). The code accounts for the full eLISA orbital motion during the observation time, but also uses the analytical approximation for the sensitivity curve. We checked that, given ${\cal M}$ and $f_i$, the S/N returned by the code matches the estimate of equation (\ref{eqsnr}) when averaged over a Monte Carlo realization of the parameters describing the source sky location, inclination and polarization.

Finally, the estimate of the stochastic signal is computed following \cite{2013PhRvD..88l4032T} as
\begin{equation}
  ({\rm S/N})^2_{\rm bkg} = T\int\gamma(f)\frac{h_{c,\rm bkg}^4(f)}{4f^2\langle S(f)\rangle^2}df,
  \label{snrbkg}
\end{equation}
where $T=5$yr is the mission lifetime and we used the fact that $h_{c,\rm bkg}^2(f)=fS_h(f)$, being $S_h(f)$ the power spectral density of the signal. Note that the response function $\gamma(f)\approx 1$ in the relevant frequency range (see figure 4 in \cite{2013PhRvD..88l4032T}). $h_{c,\rm bkg}$ is related to the GW energy density via $h_{c,\rm bkg}^2=3H_0^2\Omega_{\rm gw}^2/(2\pi^2f^2)$, and is calculated at each frequency by summing in quadrature the characteristic strains of all sources up to $z=2$. In our simple estimate we did not remove sources with individual S/N$>8$, which however do contribute to less than $10\%$ to the estimate of the background. This is compensated by the fact that we integrate up to $z=2$, whereas significant contribution to the background comes from higher redshifts. However, we cannot trust (already at $z=2$ in fact) the assumption of a constant intrinsic BHB merger rate and our stochastic background S/N estimates are only indicative.

\begin{figure}
\includegraphics[scale=0.4,clip=true,angle=0]{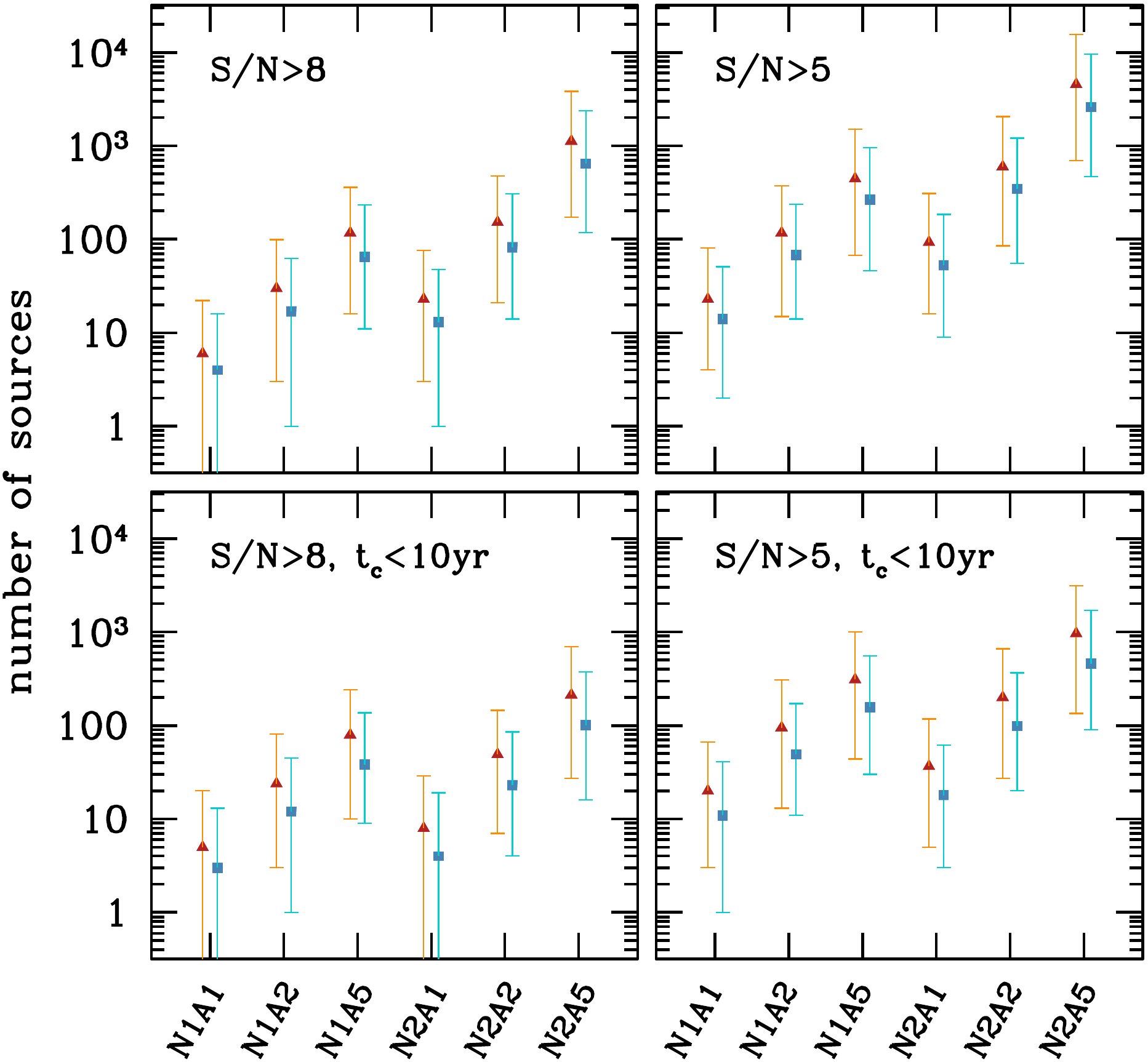}
\caption{Number of BHBs resolved by eLISA for different baselines. Orange triangles and blue squares are for models {\it flat} and {\it salp} respectively. Filled symbols and associated error-bars represent the median and 95\% confidence interval from 200 realizations of the BHB population. The two top panels represent the total number of resolved sources above the indicated threshold. The two lower panels depict the subset of sources that will eventually coalesce in the aLIGO band within 10 years from the start of the eLISA mission. All figures are computed assuming five years of eLISA operations.}
\label{fig2}
\end{figure}

\begin{figure}
\includegraphics[scale=0.4,clip=true,angle=0]{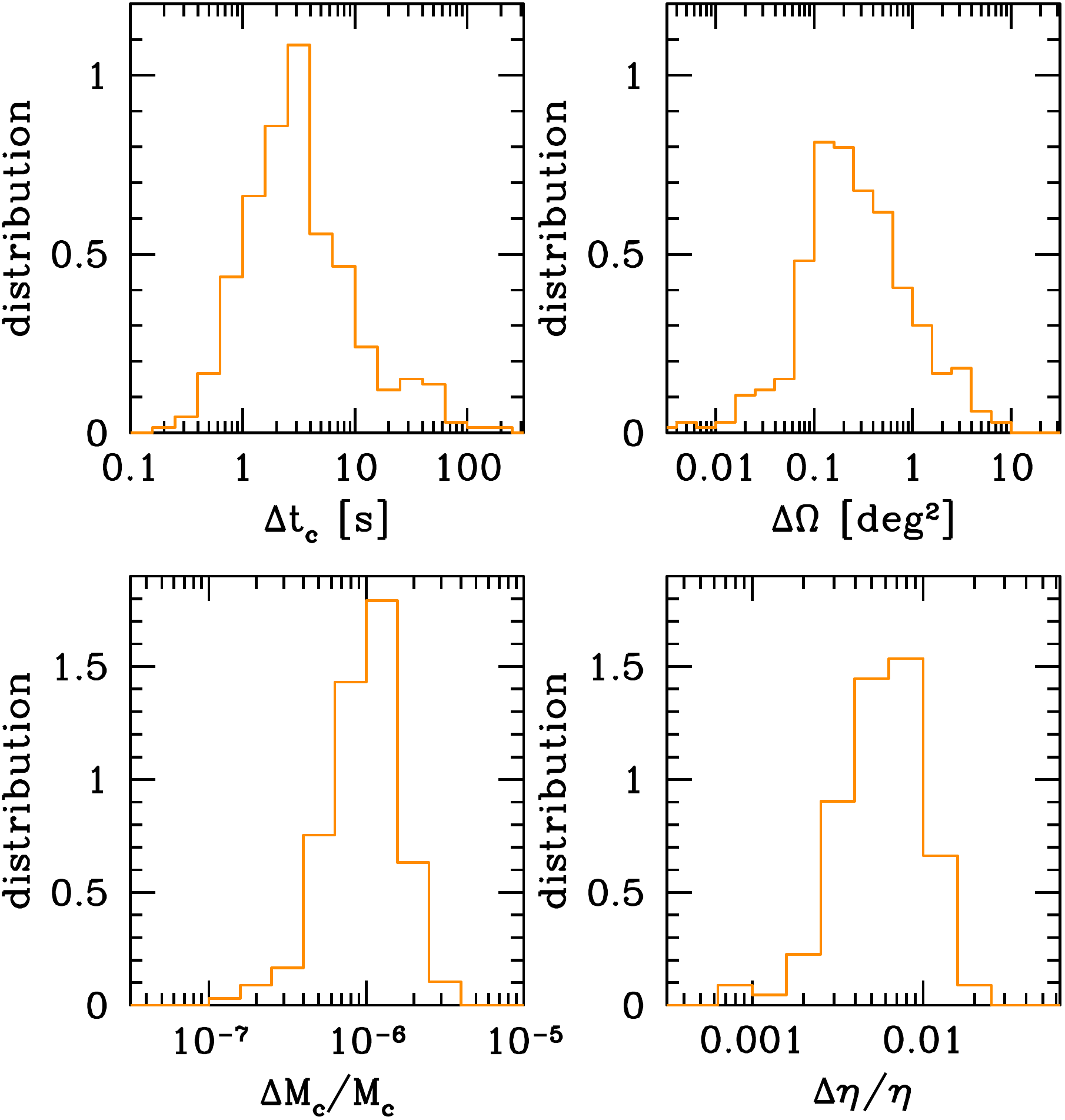}
\caption{Parameter estimation precision from eLISA observations. Top left: coalescence time; top right: sky localization; bottom left: relative error in the chirp mass ${\cal M}$; bottom right: relative error in the symmetric mass ratio ${\eta}=M_1M_2/(M_1+M_2)^2$. Histograms show normalized distributions obtained from a Monte Carlo realization of 1000 sources observed with S/N$>8$ in the N2A5 configuration for five years of mission operation. Estimates were obtained via Fisher Matrix analysis using 3.5PN non spinning waveforms \cite{2005PhRvD..71h4025B} and the full time-dependent eLISA response function.}
\label{fig3}
\end{figure}

{\it Results and implications.} For each configuration we select only events resolvable above a given signal-to-noise ratio (S/N) threshold. Results are shown in figure \ref{fig2}. Between one and about a thousand BHBs will be observable at S/N$>8$, and a factor of about four more at S/N$>5$, with the {\it flat} model resulting in twice as many sources as the {\it salp} one. Four link configurations would yield approximately one third of detections, since their sensitivity is a factor $\sqrt{2}$ smaller, and the cumulative number of sources goes with (S/N)$^3$. About 20\% of the resolvable systems will coalesce within ten years from the start of eLISA operations, appearing into the aLIGO band. These are typically massive binaries ($50\msun<M_1+M_2<100\msun$) and can be observed up to $z\approx 0.4$ in eLISA. Numbers are therefore quite sensitive to the high end of the BHB mass function, but even assuming an artificial pessimistic cut-off for systems more massive than GW150914, we obtain tens of events for the best eLISA design.

Figure \ref{fig3} shows an example of parameter estimation precision achievable with eLISA, for a typical population of systems coalescing in the aLIGO band within its lifetime. The plot was constructed by running the Fisher Matrix code on a sub-sample of 1000 sources coalescing in five years and resulting in an S/N$>8$ in the eLISA detector (configuration N2A5, but distributions are largely insensitive to the specific design), taken from our 200 Monte Carlo realizations of the {\it flat} BHB mass model. The exquisite precision is due to the many thousands of wave cycles emitted by the system convolved with the multiple orbits completed by the eLISA detector over five years. Although we use a simple waveform and detector response model, adding complexity to the waveform and to the response function should not appreciably alter the precision of the measurement, as discussed above. Typically few weeks before appearance in the aLIGO band, the relative errors in the mass measurements is better than $1\%$, the sky location is better than 1deg$^2$, and the coalescence time can be predicted within less than ten seconds. These figures open the possibility to mutually enhance the capabilities of aLIGO and eLISA and to open the era of multi-band GW astronomy.

Electromagnetic counterparts to BHB coalescences are theoretically not expected, unless matter (likely ionized hot gas in form of some accretion disk) is also present. However, a tentative gamma  signal coincident with GW150914 has been detected by the Gamma-ray Burst Monitor (GBM) on board Fermi \cite{2016arXiv160203920C}, a nearly all-sky monitor with necessarily limited sensitivity and angular resolution. The fact that no alert can be sent to satellites and telescopes {\it prior} to coalescence fundamentally limits the possibility of real-time electromagnetic observations of aLIGO BHBs by telescopes with more restricted field of view and higher sensitivity (see a review in \cite{2016arXiv160208492A}).
However, for up to a couple of hundred sources in the best configuration, eLISA can alert aLIGO and all possible electromagnetic probes weeks in advance, providing the exact location and time of the merger. Firstly, this will allow the aLIGO team to plan the operation schedule ensuring at least two interferometers will be in operation during these events, reducing the loss of events due to missing detector coincidence. Secondly, all the most sensitive telescopes covering the sky from the radio to the $\gamma$-ray, can then be pre-pointed securing the detection of a prompt counterpart at any wavelength, should there be one, opening new horizons in multimessenger astronomy. Moreover, eLISA will determine the individual masses of the two systems within $<1$\% precision, possibly constraining also their spins. This wealth of information can be used to pin-down the pre-merger properties of the BHB to a level that is unthinkable with aLIGO only, tremendously improving the feasibility of fundamental physics and strong gravity tests \cite{2014PhRvD..89h2001A,2016arXiv160203841T}. For example, \cite{2016arXiv160304075B} found that constrains on BH dipole radiation can be improved by five orders of magnitudes compared to observations with aLIGO alone. Hundreds of low redshift GW sources with accurate sky localization also make for a new interesting population of cosmological standard sirens \cite{1986Natur.323..310S} that can be exploited following the idea put forward in \cite{2008PhRvD..77d3512M} for extreme mass ratio inspirals. On the other hand, aLIGO will likely see BHB mergers that have an S/N$<8$ in the eLISA data-stream (see figure \ref{fig1}). Those can be used as triggers to search back in the eLISA data for sub-threshold signals. Equivalently, one can flag all events with a S/N much lower than the confident detection threshold in the eLISA data-stream, and wait for their aLIGO confirmation. Lastly, these systems provide a unique consistency test-bed for the two instruments, that can be the ultimate cross-band check vetting their mutual calibration.

\begin{figure}
\includegraphics[scale=0.4,clip=true,angle=0]{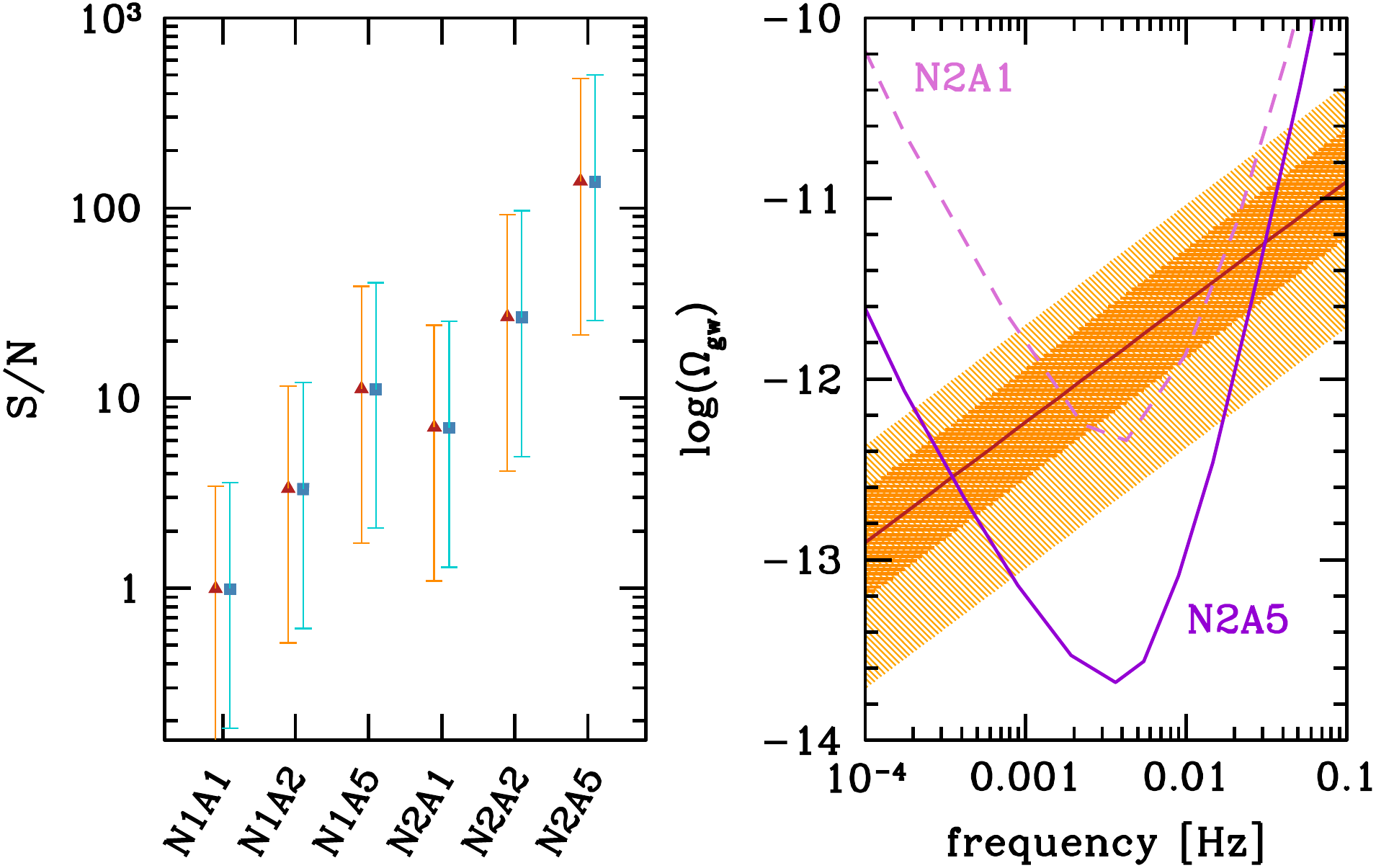}
\caption{Unresolved BHB confusion noise in the eLISA detector. The left panel shows the S/N of the unresolved confusion noise for different eLISA designs, assuming two Michelson (six links, L6) on a five year baseline. The filled triangles and associated error-bars represent the median and 95\% confidence interval of the S/N from 200 realizations of the BHB population. Orange and blue symbols are for models {\it flat} and {\it salp} respectively. The right panel shows the energy density content of the confusion noise as a function of frequency, $\Omega_{gw}(f)$ (median, 68\% and 95\% intervals are shown). This is compared to the eLISA sensitivity to a stochastic background \cite{2013PhRvD..88l4032T} for two different baselines (as indicated in figure).}
\label{fig4}
\end{figure}

Below the resolvable sources, there is an unresolved confusion noise of the same nature of the one generated by WD-WD binaries \cite{2001A&A...375..890N}. We find that this confusion noise will affect the bottom of the eLISA sensitivity curve only for optimistic BHB merger rates in combination with the best detector configuration (see figure \ref{fig1}), and therefore should note pose a serious issue for the detectability of other low S/N sources such as extreme mass ratio inspirals \cite{2009CQGra..26i4034G}. However, {\it only} in six link baselines, laser links can be combined appropriately to make the background measurement feasible \cite{1999ApJ...527..814A} even without the standard cross correlation analysis \cite{1999PhRvD..59j2001A}. The expected S/N, computed via equation (\ref{snrbkg}), is in the range $1-200$, depending on the baseline. Considering all sources to $z=10$ would increase the S/N by a mere 20\%. We caution, however, that we assumed a cosmologically non-evolving BHB merger rate. Although this might by a safe ansatz at the low redshifts relevant to the statistics of resolvable sources, it is bound to break down beyond the local Universe. To investigate the possible effect of a BHB merger rate evolving with redshift, we considered the two scenarios proposed in \cite{2016MNRAS.tmp..160M} and \cite{2016arXiv160204531B}, which are representative of two different BHB formation channels, normalizing both rates to a fiducial value of $50$yr$^{-1}$Gpc$^{-3}$ at $z=0$. Although \cite{2016MNRAS.tmp..160M} results in a GW background in line with the non-evolving coalescence rate model, \cite{2016arXiv160204531B} predicts an $\Omega_{\rm gw}(f)$ larger by a factor $\approx$3.5. Therefore, the unresolved background might provide some valuable information about the high redshift abundance of BHBs and their formation channel, complementary to low redshift observations of individual sources.


{\it Outlook.} The observation of GW150914 brings unexpected prospects in multi-band GW astronomy, providing even more compelling evidence that a milli-Hz GW observatory will not only open a new window on the Universe, but will also naturally complete and enhance the payouts of the high frequency window probed by aLIGO. The scientific potential of multi-band GW astronomy is enormous, ranging from multimessenger astronomy, cosmology and ultra precise gravity tests with BHBs, to the study of the cosmological BHB merger rate, and to the mutual validation of the calibration of the two GW instruments. This is a unique new opportunity for the future of GW astronomy, and how much of this potential will be realized in practice, depends on the choice of the eLISA baseline. Should an extremely de-scoped design like the New Gravitational Observatory (NGO) \cite{2013GWN.....6....4A} be adopted, all the spectacular scientific prospects outlined above will likely be lost. Re-introducing the third arm (i.e. six laser links) and increasing the arm-length to at least two million kilometres (A2) will allow observation of more than 50 resolved BHB with both eLISA and aLIGO, and the detection of the unresolved confusion noise with S/N$>30$. We also stress that the most interesting systems emit at $f>10^{-2}$Hz, a band essentially 'clean' from other sources. There, the eLISA sensitivity critically depends on the shot noise, which is determined by the number of photons collected at the detector mirrors. It is therefore important to reconsider the designed mirror size and laser power under the novel appealing prospect of observing more of these BHBs and with an higher S/N.   
 
{\it Acknowledgements.} The author thanks E. Berti for the original version of the PN code used for the parameter estimation, W. Farr for carefully reading the manuscript, W. Del Pozzo, J. Gair \& A. Vecchio for useful discussions, and the anonymous referees for the valuable feedback. This work is supported by the Royal Society.

\footnotesize{
\bibliography{ligobh}
}

\end{document}